\documentclass[runningheads]{llncs}
\usepackage{graphicx}
\usepackage{hyperref}
\usepackage{fancyhdr}
\fancypagestyle{personalcopy}{\fancyhf{}\fancyfoot[L]{\footnotesize This is a personal copy of the authors.}}

\begin{document}
\title{Machine Learning Approaches to \\ Hybrid Music Recommender Systems}
%
%
\author{Andreu Vall\inst{1} \and
Gerhard Widmer\inst{1,2}}
\authorrunning{A. Vall and G. Widmer}
%
\institute{Institute of Computational Perception, Johannes Kepler University, Linz, Austria \and
Austrian Research Institute for Artificial Intelligence, Vienna, Austria
\email{\{andreu.vall,gerhard.widmer\}@jku.at}}
\maketitle              
\begin{abstract}
Music recommender systems have become a key technology supporting the access to increasingly larger music catalogs in on-line music streaming services, on-line music shops, and private collections. The interaction of users with large music catalogs is a complex phenomenon researched from different disciplines. We survey our works investigating the machine learning and data mining aspects of hybrid music recommender systems (i.e., systems that integrate different recommendation techniques). We proposed hybrid music recommender systems based solely on data and robust to the so-called ``cold-start problem'' for new music items, favoring the discovery of relevant but non-popular music. We thoroughly studied the specific task of music playlist continuation, by analyzing fundamental playlist characteristics, song feature representations, and the relationship between playlists and the songs therein.
\keywords{Music recommender systems  \and Music playlist continuation \and Hybrid recommender systems \and Cold-start problem.}
\end{abstract}
\thispagestyle{personalcopy}
\section{Introduction}

Music recommender systems support the interaction of users with large music catalogs. They strongly rely on machine learning and data mining methods to analyze the data describing the users, the music items, the interaction of users with music items, and even the users' context when interacting with the music items~\cite[Chap.~7]{ricci_recommender_2015}. Even though the techniques at the core of recommender systems are valid for different item domains (e.g., recommending movies, news, or jobs), the inherent characteristics of the music domain must be considered:
\begin{itemize}
    \item Music can be recommended at different granularity levels (e.g., songs, albums, artists, or even genres or ready-made playlists).
    \item Music is often consumed in listening sessions, such as albums or playlists. This defines local session contexts that need to be taken into consideration. 
    \item Music recommender systems should adapt their level of interference to the user needs, ranging from simply supporting the exploration of the music catalog, to providing a full \emph{lean-back} experience.
\end{itemize}

Music recommender systems have received contributions from different, but converging approaches. Research specializing in music information retrieval has often focused on content-based music recommender systems. Music items are represented by features derived from the audio signal, social tags, or web content and recommendations are predicted on the basis of content-wise similarities~\cite{flexer_playlist_2008,knees_combining_2006}. On the other hand, research specializing in general recommender systems has usually focused on collaborative filtering techniques, developing statistical models to extract underlying music taste patterns from usage data (e.g., listening logs, or radio stations)~\cite{aizenberg_build_2012,chen_playlist_2012}.

Content-based recommender systems offer a fair but limited performance, because they only identify simple relations derived from content-wise similarities. Collaborative recommender systems capture more abstract music taste patterns, but their performance is heavily affected by the availability of data (users and music items for which few observations are available are poorly represented).

In this paper we survey our recent works on hybrid music recommender systems integrating the strengths of content-based and collaborative recommender systems. We approached two main music recommendation tasks: 1) music artist recommendation, modeling the users' general music preferences over an extended period of time, and 2) music playlist continuation, focusing on next-song recommendations for short listening sessions.

\section{Music Artist Recommendation}

We proposed in~\cite{vall_improving_2015} a hybrid extension to a well-established matrix factorization CF approach for implicit feedback datasets~\cite{hu_collaborative_2008}. The proposed hybrid CF extension jointly factorizes listening logs and social tags from music streaming services (abundant sources of usage data in the music domain). According to our numerical experiments, the proposed hybrid CF extension yielded more accurate music artist recommendations. Furthermore, we extended the standard evaluation methodology incorporating Bootstrap confidence intervals~\cite{diciccio_bootstrap_1996} to facilitate the comparison between systems. In the follow-up work~\cite{vall_listener-inspired_2015} we observed that the superior performance of the proposed hybrid CF extension was explained by its robustness to the cold-start problem for new artists (i.e., its ability to better represent music artists for which few observations were available).

\section{Music Playlist Continuation}

Like previous works on music playlist modeling, we based our research on the exploitation of hand-curated music playlists, which we regard as rich examples from which to learn music compilation patterns.

\subsection{Playlist Characteristics and Next-Song Recommendations}

Before moving into the design of hybrid music recommender systems for music playlist continuation, we studied which basic playlist characteristics should be considered to effectively predict next-song recommendations. We studied in~\cite{vall_importance_2018,vall_importance_2017} the importance of the song order, the song context, and the song popularity for next-song recommendations. We compared four existing playlist continuation models of increasing complexity on two datasets of hand-curated music playlists. We observed that considering a longer song context has a positive impact on next-song recommendations. We found that the long-tailed nature of the playlist datasets~(common in music collections~\cite{celma_music_2010}) makes simple and highly-expressive playlist continuation models appear to perform comparably. However, further analysis revealed the advantage of using highly-expressive models. Our experiments also suggested either that the song order is not crucial for next-song recommendations, or that even highly-expressive models are unable to exploit~it. We also proposed an evaluation approach for next-song recommendations that mitigates known issues in the evaluation of music recommender systems.

\subsection{Song Representations and Playlist-Song Membership}

Given the results of the experiments described in~\cite{vall_importance_2018,vall_importance_2017}, we proposed music recommender systems for playlist continuation able to consider the full playlist song context. We also required the proposed systems to hybridize collaborative and content-based recommender systems to ensure robustness to the cold-start problem.

We identified in~\cite{vall_timbral_2016,vall_music_2017} suitable song-level feature representations for music playlist modeling. We investigated features derived from the audio signal, social tags, and independent listening logs. We found that the features derived from independent listening logs are more expressive than those derived from social tags, which in turn outperform those derived from the audio signal. The combination of features from different modalities outperformed the individual features, suggesting that the different modalities indeed carry complementary information.

We further proposed in~\cite{vall_music_2017} a hybrid music recommender system for music playlist continuation robust to the cold-start problem for non-popular songs. However, this approach can only extend playlists for which a profile has been pre-computed at training time. In the follow-up work~\cite{vall_hybrid_2018} we proposed another hybrid music recommender system for music playlist continuation that regards playlist-song pairs exclusively in terms of feature vectors. This system learns general ``playlist-song'' membership relationships, which not only make it robust to the cold-start problem for non-popular songs but also enable the extension of playlists not seen at training time.

\section{Lessons Learned and Open Challenges}

Additional insights and questions arise from the research conducted in the presented works. Importantly, the evaluation of music recommender systems by means of numerical experiments does not reflect the fact that the usefulness of music recommendations is a highly subjective judgment of the end user. Further research on evaluation metrics for music recommender systems is required. On a related note, the users' subjectivity on the usefulness of music recommendations makes it challenging to anticipate the actual immediate user needs. An interesting research direction to bridge this gap focuses on the development of interactive interfaces that should let users express their current needs.

\paragraph{Acknowledgments.}

This research has received funding from the European Research Council (ERC) under the European Union's Horizon 2020 research and innovation programme under grant agreement No 670035 (Con Espressione).

%
%
%
\bibliographystyle{splncs04}
\bibliography{nectar2018}
\end{document}